# Branching of the $W(H_4)$ Polytopes and Their Dual Polytopes under the Coxeter Groups $W(A_4)$ and $W(H_3)$ Represented by Quaternions

**Mehmet Koca [1], Nazife Ozdes Koca [2]**, and **Mudhahir Al-Ajmi [3]**

[1, 2, 3] Department of Physics, College of Science, Sultan Qaboos University
P. O. Box 36, Al-Khoud 123, Muscat, Sultanate of Oman

E-mail: [1]kocam@squ.edu.om, [2] nazife@squ.edu.om, [3]mudhahir@squ.edu.om

**Abstract.** 4-dimensional $H_4$ polytopes and their dual polytopes have been constructed as the orbits of the Coxeter-Weyl group $W(H_4)$ where the group elements and the vertices of the polytopes are represented by quaternions. Projection of an arbitrary $W(H_4)$ orbit into three dimensions is made preserving the icosahedral subgroup $W(H_3)$ and the tetrahedral subgroup $W(A_3)$, the latter follows a branching under the Coxeter group $W(A_4)$. The dual polytopes of the semi-regular and quasi-regular $H_4$ polytopes have been constructed.

## 1. Introduction

It seems that there exists an experimental evidence for the realization of the Coxeter-Weyl group $W(E_8)$. Radu Coldea from the University of Oxford and his colleagues, from Oxford, Bristol University, the ISIS Rutherford Laboratory and the Helmholtz Zentrum Berlin [1] performed a neutron scattering experiment on $CoNb_2O_6$ (cobalt niobate) which describes the one dimensional quantum Ising chain. They have determined masses of the five emerging particles, first two are obeying the relation $m_2 = \tau m_1$. Their results could be attributed to the Zamolodchikov model [2] which describes the one-dimensional Ising model at critical temperature perturbed by an external magnetic field leading to eight spinless bosons with the mass relations

$$m_1, \qquad m_3 = 2m_1 \cos\frac{\pi}{30}, \quad m_4 = 2m_2 \cos\frac{7\pi}{30}, \quad m_5 = 2m_2 \cos\frac{4\pi}{30}$$

$$m_2 = \tau m_1, \quad m_6 = \tau m_3, \qquad m_7 = \tau m_4, \qquad m_8 = \tau m_5 \qquad (1)$$

where $\tau = \frac{1+\sqrt{5}}{2}$ is the golden ratio. Zamolodchikov model can be described by an affine Toda Field Theory of $E_8$ [3]. The masses can be related to the radii of the Gosset's polytope of $E_8$ [4]. In the derivation of the mass relations in (1) the maximal subgroup $W(H_4)$ [5] of the Coxeter-Weyl group $W(E_8)$ plays a crucial role. This is evident from the relation (1) where four of the masses are the $\tau$ times the other four masses. This motivates us to study the structure of the group $W(H_4)$ in terms of quaternions and its 4D polytopes.

The rank-4 Coxeter group $W(H_4)$ is unique in the sense that it has no correspondence in higher dimensions and describes the quasi crystallography in 4-dimensions. It is the extension of the icosahedral group $W(H_3) \approx A_5 \times C_2 \approx I_h$ to 4-dimensions. The group $W(H_4)$ has four 4-dimensional irreducible representations. One of its 4-dimensional irreducible representations can be described by the transformations on the quaternions by performing a left-right multiplication of the binary icosahedral group elements $I$ which constitute one of the finite subgroups of quaternions. The Coxeter group $W(H_4)$ has five maximal subgroups [6] described by the groups

$$Aut(A_4) \approx W(A_4):C_2, \ W(H_3 \oplus A_1), \ Aut(H_2 \oplus H_2), \ Aut(A_2 \oplus A_2), \ (\frac{W(D_4)}{C_2}):S_3. \qquad (2)$$

We use the notation (:) for the semi direct product of two groups. Importance of these groups will be discussed in the next section.

It is perhaps interesting to note that the Coxeter-Weyl group $W(E_8)$ seems to show itself as an affine Toda field theory in the zero temperature (the coldest regime) and perhaps as a Lie group in the form of $E_8 \times E_8$ describing the heterotic superstring theory at very hot regime (Planck scale). The Coxeter-Weyl group $W(E_8)$ includes the crystallographic (tetrahedral and octahedral symmetries in 3D and 4D) as well as the quasi crystallographic symmetries (icosahedral symmetry in 3D and its generalization to 4D). The icosahedral group of rank-3 describes fully the structures of the fullerenes such as the $C_{60}$ molecule which is represented by a truncated icosahedron, the icosahedral quasicrystals and the viral structures displaying the icosahedral symmetry.

A short historical account of the development of the theory of polytopes could be in order. The Platonic solids, tetrahedron, cube, octahedron, icosahedron and dodecahedron have been discovered by the people lived in Scotland nearly 1000 years earlier than the ancient Greeks. Their models curved on the stones are now kept in the Ashmolean Museum at Oxford [7]. They were used in old Greek as models to describe fundamental matter associating tetrahedron with fire, cube with earth, air with octahedron, and water with icosahedron and dodecahedron with the universe or ether. Archimedes discovered the semi-regular convex solids now known as Archimedean solids. Kepler completed the classifications of the regular polyhedra in 1620 by inventing prisms and anti-prisms as well as four regular non-convex polyhedra. In 1865 the Belgian mathematician Catalan constructed the dual solids of the Archimedean solids, now known as Catalan solids [8]. Extensions of the platonic solids to 4D dimensions have been made in 1855 by L. Schlaffli [9] and their generalizations to higher dimensions in 1900 by T. Gosset [10]. Further important contributions are made by W. A. Wythoff [11] among many others and, in particular, by the contemporary mathematicians H.S.M. Coxeter [12] and J.H. Conway [13].

In what follows we study the branching of the regular, semi regular and quasi regular 4D polytopes described as the orbits of the Coxeter group $W(H_4)$. Constructions of the defining representation of the Coxeter group $W(H_4)$ and its maximal subgroups in terms of quaternions provide an appropriate technique to pursue. For this reason we introduce the finite subgroups of quaternions in Section 2 and construct the group elements of the groups $W(H_4)$, $W(A_4)$ and $W(H_3)$ in terms of quaternions. The Section 3 deals with the decomposition of the $W(H_4)$ polytopes under the subgroup $W(H_3)$. A similar analysis is carried in Section 4 for the decomposition under the subgroup $W(A_4)$. Noting that the tetrahedral group $W(A_3)$ is a subgroup of the group $W(A_4)$ we give a few examples of the projections of the $W(H_4)$ polytopes into 3D preserving the tetrahedral symmetry. Dual polytopes of the 4D semi regular polytopes have not been studied in mathematical literature. The section 5 deals with the construction of the dual polytopes of the regular and semi regular polytopes of the Coxeter group $W(H_4)$. Some concluding remarks are made in Section 6.

**2. Finite subgroups of quaternions**

This section deals with the quaternions and its relevance to the orthogonal transformations in 4- dimensions and gives the list of finite subgroups of quaternions.

*2.1. Quaternions and $O(4)$ Transformations*

Let $q = q_0 + q_i e_i$, $(i = 1, 2, 3)$, be a real unit quaternion with its conjugate defined by $\bar{q} = q_0 - q_i e_i$ and the norm $q\bar{q} = \bar{q}q = 1$. The imaginary units satisfy

$$e_i e_j = -\delta_{ij} + \varepsilon_{ijk} e_k, \quad (i, j, k = 1, 2, 3). \tag{3}$$

Let $p, q$ be unit quaternions and $r$ represents an arbitrary quaternion. Then the transformations [14]

$$r \to prq : [p, q]; \quad r \to p\bar{r}q : [p, q]^* \tag{4}$$

define the orthogonal group $O(4)$ which preserves the norm $r\bar{r} = \bar{r}r$. We shall use the abstract notations $[p, q]$ and $[p, q]^*$ respectively for the transformations in (4). The first term in (3) represents a proper rotation and the second includes also the reflection, generally called rotary reflection. In particular, the group element

$$r \to -p\bar{r}p : [p, -p]^* \tag{5}$$

represents the reflection with respect to the hyperplane orthogonal to the unit quaternion $p$.

The orthogonal transformations in 3D can be simply written as $r \to \pm p r \bar{p} : \pm[p, \bar{p}]$ where $r$ is an imaginary quaternion.

## 2.2. Finite Subgroups of Quaternions

The finite subgroups of quaternions are well known and its classification can be found in the references [15]. They are given as follows.

(a) Cyclic group of order $n$ with $n$ an odd number.

(b) Cyclic group of order $2n$ is generated by $\langle p = \exp(e_1 \frac{\pi}{n}) \rangle$ and dicyclic group of order $4n$ can be generated by the generators $\langle p = \exp(e_1 \frac{\pi}{n}), e_2 \rangle$.

(c) The binary tetrahedral group can be represented by the set of 24 unit quaternions:

$$T = \{ \pm 1, \pm e_1, \pm e_2, \pm e_3, \frac{1}{2}(\pm 1 \pm e_1 \pm e_2 \pm e_3) \}. \tag{6}$$

The set $T$ represents also the vertices of the 24-cell.

(d) The binary octahedral group consists of 48 unit quaternions. Let the set

$$T' = \{\tfrac{1}{\sqrt{2}}(\pm 1 \pm e_1), \tfrac{1}{\sqrt{2}}(\pm e_2 \pm e_3), \tfrac{1}{\sqrt{2}}(\pm 1 \pm e_2), \tfrac{1}{\sqrt{2}}(\pm e_3 \pm e_1), \tfrac{1}{\sqrt{2}}(\pm 1 \pm e_3), \tfrac{1}{\sqrt{2}}(\pm e_1 \pm e_2)\} \tag{7}$$

represents the vertices of the 24-cell rotated with respect to the set $T$ [16].

The union of the set $O = T \oplus T'$ represents the binary octahedral group of order 48.

(e) The binary icosahedral group $I = \langle b, c \rangle$ of order 120 can be generated by two unit quaternions, say, $b = \frac{1}{2}(\tau + \sigma e_1 + e_2) \in 12(1)_+$ and $c = \frac{1}{2}(\tau - \sigma e_1 + e_2) \in 12(1)_+$ belonging to the same conjugacy classes. We display its elements in Table 1 as the sets of conjugacy classes which represent a number of icosahedra, dodecahedra and one icosidodecahedron.

**Table 1**. Conjugacy classes of the binary icosahedral group $I$ represented as the orbits of the group $W(A_3)$.

| | Conjugacy classes denoted by the number of elements and order of elements |
|---|---|
| 1 | 1 |
| 2 | -1 |
| 10 | $12_+ : \frac{1}{2}(\tau \pm e_1 \pm \sigma e_3), \frac{1}{2}(\tau \pm e_2 \pm \sigma e_1), \frac{1}{2}(\tau \pm e_3 \pm \sigma e_2)$ |
| 5 | $12_- : \frac{1}{2}(-\tau \pm e_1 \pm \sigma e_3), \frac{1}{2}(-\tau \pm e_2 \pm \sigma e_1), \frac{1}{2}(-\tau \pm e_3 \pm \sigma e_2)$ |
| 10 | $12'_+ : \frac{1}{2}(\sigma \pm e_1 \pm \tau e_2), \frac{1}{2}(\sigma \pm e_2 \pm \tau e_3), \frac{1}{2}(\sigma \pm e_3 \pm \tau e_1)$ |
| 5 | $12'_- : \frac{1}{2}(-\sigma \pm e_1 \pm \tau e_2), \frac{1}{2}(-\sigma \pm e_2 \pm \tau e_3), \frac{1}{2}(-\sigma \pm e_3 \pm \tau e_1)$ |
| 6 | $20_+ : \frac{1}{2}(1 \pm e_1 \pm e_2 \pm e_3), \frac{1}{2}(1 \pm \tau e_1 \pm \sigma e_2), \frac{1}{2}(1 \pm \tau e_2 \pm \sigma e_3), \frac{1}{2}(1 \pm \tau e_3 \pm \sigma e_1)$ |
| 3 | $20_- : \frac{1}{2}(-1 \pm e_1 \pm e_2 \pm e_3), \frac{1}{2}(-1 \pm \tau e_1 \pm \sigma e_2), \frac{1}{2}(-1 \pm \tau e_2 \pm \sigma e_3), \frac{1}{2}(-1 \pm \tau e_3 \pm \sigma e_1)$ |
| 4 | $30 : \pm e_1, \pm e_2, \pm e_3, \frac{1}{2}(\pm \sigma e_1 \pm \tau e_2 \pm e_3), \frac{1}{2}(\pm \sigma e_2 \pm \tau e_3 \pm e_1), \frac{1}{2}(\pm \sigma e_3 \pm \tau e_1 \pm e_2)$ |

In Table 1 the golden ratio $\tau = \frac{1+\sqrt{5}}{2}$ and $\sigma = \frac{1-\sqrt{5}}{2}$ satisfy the relations $\tau + \sigma = 1$, $\tau\sigma = -1$, $\tau^2 = \tau + 1$, $\sigma^2 = \sigma + 1$. Note that the 30 quaternions in the last line of Table 1 consist of imaginary quaternions and constitute the vertices of an icosidodecahedron.

The set of quaternions representing the binary icosahedral group $I$ can also be written as the union of two sets,

$$I = \{T \oplus S\} \quad (8)$$

where the set $S$ is represented by the quaternions

$$S = \{\frac{1}{2}(\pm\tau \pm e_1 \pm \sigma e_3), \frac{1}{2}(\pm\tau \pm e_2 \pm \sigma e_1), \frac{1}{2}(\pm\tau \pm e_3 \pm \sigma e_2),$$

$$\frac{1}{2}(\pm\sigma \pm e_1 \pm \tau e_2), \frac{1}{2}(\pm\sigma \pm e_2 \pm \tau e_3), \frac{1}{2}(\pm\sigma \pm e_3 \pm \tau e_1),$$

$$\frac{1}{2}(\pm 1 \pm \tau e_1 \pm \sigma e_2), \frac{1}{2}(\pm 1 \pm \tau e_2 \pm \sigma e_3), \frac{1}{2}(\pm 1 \pm \tau e_3 \pm \sigma e_1),$$

$$\frac{1}{2}(\pm\sigma e_1 \pm \tau e_2 \pm e_3), \frac{1}{2}(\pm\sigma e_2 \pm \tau e_3 \pm e_1), \frac{1}{2}(\pm\sigma e_3 \pm \tau e_1 \pm e_2)\}. \tag{9}$$

It is interesting to note that the set of quaternions $I$ represents the vertices of 600-cell [17] and the set $S$ represents the vertices of the snub 24-cell [18].

The Coxeter diagram $H_4$ is illustrated in Figure 1 where the simple roots are given by the quaternions

$$\alpha_1 = -\sqrt{2}e_1, \quad \alpha_2 = \frac{\sqrt{2}}{2}(\tau e_1 + e_2 + \sigma e_3), \quad \alpha_3 = -\sqrt{2}e_2, \quad \alpha_4 = \frac{\sqrt{2}}{2}(\sigma + e_2 + \tau e_3) \tag{10}$$

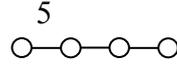

**Figure 1.** The Coxeter diagram of $H_4$.

Let $\omega_i$ $(i = 1,2,3,4)$ be the basis vectors of the dual space defined by $(\omega_i, \alpha_j) = \delta_{ij}$. The Cartan matrix with the matrix elements $(C)_{ij} = (\alpha_i, \alpha_j)$ and its inverse with the matrix elements $(C^{-1})_{ij} = (\omega_i, \omega_j)$ are given by the respective matrices

$$C_{H_4} = \begin{pmatrix} 2 & -\tau & 0 & 0 \\ -\tau & 2 & -1 & 0 \\ 0 & -1 & 2 & -1 \\ 0 & 0 & -1 & 2 \end{pmatrix}, \quad (C^{-1})_{H_4} = \tau^4 \begin{pmatrix} 4 & 3\tau & 2\tau & \tau \\ 3\tau & 6 & 4 & 2 \\ 2\tau & 4 & 2(2+\sigma) & 2+\sigma \\ \tau & 2 & 2+\sigma & 2\sigma^2 \end{pmatrix}. \tag{11}$$

Note that the relations $\alpha_i = (C_{H_4})_{ij} \omega_j$ and $\omega_i = (C_{H_4}^{-1})_{ij} \alpha_j$ lead to the quaternionic representations of the dual vectors

$$\omega_1 = -\frac{1}{\sqrt{2}}(\tau^4 + e_1 + \tau^2 e_3), \qquad \omega_2 = -\sqrt{2}(\tau^3 + \tau e_3),$$

$$\omega_3 = -\frac{1}{\sqrt{2}}(\tau(\tau+2) + e_2 + \tau e_3), \quad \omega_4 = -\sqrt{2}\tau. \tag{12}$$

Following equation (5) we define the generators of the group $W(H_4) = \langle r_1, r_2, r_3, r_4 \rangle$ by $r_i = \frac{1}{2}[\alpha_i, -\alpha_i]^*$ which leads to the quaternionic representation of the Coxeter group $W(H_4) = [p,q] \oplus [p,q]^*$ of order 14,400 where $p, q \in I$. We can use a symbolic notation $W(H_4) = [I,I] \oplus [I,I]^*$ for the designation of the group. Now the subgroup $W(H_3)$ can be represented by the set $W(H_3) = [p, \bar{p}] \oplus [p, \bar{p}]^*$ with $p, \bar{p} \in I$. It is obvious that the group can be represented by the set $W(H_3) = [I, \bar{I}] \oplus [I, \bar{I}]^* \approx A_5 \times C_2$ where $A_5$ is the group of even permutations of five letters with an order of 60 and the $C_2$ stands for the group generated by inversion. An extended $H_4$ diagram allows us to represent an extended root represented by the quaternion $\alpha_0 = \sqrt{2}$. When the root $\alpha_4$ is deleted the remaining diagram represents the group structure $W(H_3 \oplus A_1) = \langle r_1, r_2, r_3 \rangle \times \langle r_0 \rangle$ where the generator $r_0$ can be represented by reflection $r_0 = [1, -1]*$ with respect to the hyperplane orthogonal to the unit quaternion 1. Then the representation of the maximal subgroup of order 240 is given by

$$W(H_3) \times W(A_1) = [I, \pm \bar{I}] \oplus [I, \pm \bar{I}]^* \approx A_5 \times C_2 \times C_2. \quad (13)$$

Since the index $\frac{|W(H_4)|}{|W(H_3) \times W(A_1)|} = 60$ then the maximal subgroup $W(H_3) \times W(A_1)$ can be embedded in the Coxeter group $W(H_4)$ in 60 different ways which are all conjugate to each other. The group in (13) leaves the unit quaternion $\pm 1$ invariant. The other conjugate groups can be represented by the set

$$\{W(H_3) \times W(A_1)\}^q = [I, \pm \bar{q} \bar{I} q] \oplus [I, \pm q \bar{I} q]^* \quad (14)$$

which leaves $\pm q$ invariant where $q \in I$ is an arbitrary element of the binary icosahedral group $I$.

The subgroup $W(A_4)$ can be generated by the reflections on the simple roots of the Weyl-Coxeter diagram shown in Fig. 2.

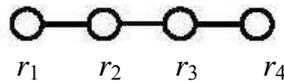

**Figure 2.** The Coxeter diagram of $A_4$.

One possible choice of the simple roots can be taken from the set $I$ as follows

$$\alpha_1 = -\sqrt{2}, \quad \alpha_2 = \frac{\sqrt{2}}{2}(1 + e_1 + e_2 + e_3), \quad \alpha_3 = -\sqrt{2} e_1, \quad \alpha_4 = \frac{\sqrt{2}}{2}(e_1 - \sigma e_2 - \tau e_3). \quad (15)$$

The Coxeter-Weyl group $W(A_4) = \langle r_1, r_2, r_3, r_4 \rangle$ then takes the form [6] $W(A_4) = \{[p, c\bar{p}c] \oplus [p, c\bar{p}c]^*\} \approx S_5 \approx A_5 : C_2$ where $p \in I$ with $\tilde{p} = p(\tau \leftrightarrow \sigma)$ is an element of the binary icosahedral group $\tilde{I}$ obtained from $I$ by interchanging $\tau$ and $\sigma$ and $c = \frac{1}{\sqrt{2}}(e_3 - e_2) \in T'$ is a fixed quaternion. The Dynkin diagram symmetry $\gamma : \alpha_1 \leftrightarrow \alpha_4, \alpha_2 \leftrightarrow \alpha_3$ can be represented by the quaternions $\gamma = [a, b]$ with $a = \frac{1}{2}(-\tau e_1 + e_2 + \sigma e_3)$ and $b = \frac{1}{2}(\sigma e_1 - \tau e_2 - e_3)$ leading to the group extension, namely,

$$Aut(A_4) = W(A_4) : C_2 = \{[I, \pm \bar{c}\, \bar{\tilde{I}}\, c] \oplus [I, \pm c\, \bar{\tilde{I}}\, c]^*\} \approx (A_5 : C_2) : C_2 . \qquad (16)$$

The group $Aut(A_4)$ is of the order 240 and can be embedded in the Coxeter group $W(H_4)$ in 60 different ways one for each $\pm q \in I$ pairs. Before we conclude this section we also mention that another 4-dimensional irreducible representation of the group $W(H_4)$ can be described as $W(H_4) = [\tilde{I}, \tilde{I}] \oplus [\tilde{I}, \tilde{I}]^*$. Then the above analysis can be carried out with the interchange of two irreducible representations of the binary icosahedral group $I \leftrightarrow \tilde{I}$. The other maximal subgroups of the Coxeter group $W(H_4)$ are also of special interest. The groups $Aut(H_2 \oplus H_2)$ and $(\frac{W(D_4)}{C_2}) : S_3$ are the symmetry groups of the Grand Antiprism and the snub 24-cell respectively. The group $Aut(A_2 \oplus A_2) \approx [W(A_2) \times W(A_2)] : C_4$ has some applications in particle physics where the group $W(A_2) \times W(A_2)$ is the skeleton of the chiral Lie group $SU(3)_L \times SU(3)_R$.

## 3. The $W(H_4)$ polytopes and branching under the group $W(H_3)$

An arbitrary 'highest' weight vector $\Lambda$ describing an irreducible representation of a rank-4 Lie group is given in the Dynkin basis [19] as $\Lambda = (a_1 a_2 a_3 a_4) \equiv a_1 \omega_1 + a_2 \omega_2 + a_3 \omega_3 + a_4 \omega_4$ where $\omega_i, (i = 1,2,3,4)$ are the vectors of the dual basis satisfying the relation $(\alpha_i, \omega_j) = \delta_{ij}$. For the 'highest' weight vector the Dynkin indices are non-negative integers $a_i \geq 0$. Although the Coxeter group $W(H_4)$ is not associated with any Lie group one can still use the same technique to determine the orbits of a given vector. Let then $\Lambda = (a_1 a_2 a_3 a_4) \equiv a_1 \omega_1 + a_2 \omega_2 + a_3 \omega_3 + a_4 \omega_4$ be a vector in the 4-dimensions with $a_i \geq 0$ any non-negative real number, not necessarily restricted to integers only.

We define an orbit of the group $W(H_4)$ by the group action $O(\Lambda) \equiv W(H_4)\Lambda$. The size of the orbit $O(\Lambda) = W(H_4)\Lambda$ is 14,400 with $a_i \neq 0$, $i = 1,2,3,4$ which is equal to the order of the group $W(H_4)$. One can count the number of cells of the polytope $O(\Lambda) = W(H_4)\Lambda$ by using its rank-3 subgroups $W(H_3)$, $W(A_3)$, $D_5 \times C_2$ and $D_3 \times C_2$. The last two groups are the symmetries of the pentagonal and triangular prisms respectively. In principle the $W(H_4)$ polytopes can be projected in 3-dimensions by preserving any one of these rank-3 subgroups. We will give examples only for the case of the groups $W(H_3)$ and $W(A_3)$. A right-coset decomposition of the group $W(H_4)$ under its subgroup $W(H_3)$ can be given by

$$W(H_4) = \sum_{i=1}^{120} \oplus\, W(H_3) g^i. \tag{17}$$

Then the orbit $O(\Lambda) = W(H_4)\Lambda$ decomposes as $O(\Lambda) = \sum_{i=1}^{120} \oplus\, W(H_3) g^i \Lambda$. Here $g^0 = [1,1]$ is the unit element. Let us define the vectors $\Lambda^{(i)} \equiv g^i \Lambda,\ i = 1, 2, ..., 120$. It is clear from (14) that the set of coset representatives $g^i$ can be chosen from the set of group elements $[I,1]$. Indeed the set $[I,1]$ form an invariant subgroup of the group $W(H_4)$. Thus the set of vectors $\Lambda^{(i)} \equiv g^i \Lambda$ are the set of 120 quaternionic vectors $I\Lambda$. Applying the group $W(H_3)$ on these 120 vectors one generates 120 different orbits each consisting of 120 vectors in general.

The group $W(H_3)$ acts on an arbitrary vector $(b_1 b_2 b_3) \equiv b_1 v_1 + b_2 v_2 + b_3 v_3$ where $(v_i, \alpha_j) = \delta_{ij}$ with $i, j = 1, 2, 3$ and the Cartan matrix $(C)_{ij} = (\alpha_i, \alpha_j)$ and its inverse $(C^{-1})_{ij} = (v_i, v_j)$ of the group $W(H_3)$ are given by

$$C_{H_3} = \begin{pmatrix} 2 & -\tau & 0 \\ -\tau & 2 & -1 \\ 0 & -1 & 2 \end{pmatrix}, \qquad (C^{-1})_{H_3} = \frac{1}{2}\begin{pmatrix} 3\tau^2 & 2\tau^3 & \tau^3 \\ 2\tau^3 & 4\tau^2 & 2\tau^2 \\ \tau^3 & 2\tau^2 & \tau+2 \end{pmatrix}. \tag{18}$$

We have to express the $W(H_3)$ orbits in the basis of the vectors $v_i$ using the relations given in (19)

$$\omega_1 = v_1 + \frac{\tau^3}{2}\omega_4,\ \omega_2 = v_2 + \tau^2 \omega_4,\ \omega_3 = v_3 + \frac{\tau+2}{2}\omega_4. \tag{19}$$

Any vector $\Lambda^{(i)} \equiv g^i \Lambda$ is now transformed to the form $(b_1, b_2, b_3, b_4) \equiv b_1 v_1 + b_2 v_2 + b_3 v_3 + b_4 \omega_4$. First three components of these do not in general satisfy the relation $b_1 \geq 0, b_2 \geq 0, b_3 \geq 0$. However there is always a vector in the orbit $W(H_3)\Lambda^{(i)}$ where the relations $b_1 \geq 0, b_2 \geq 0, b_3 \geq 0$ are satisfied. Each orbit $W(H_3)\Lambda^{(i)}$ defines a set of discrete points on a sphere $S^2$ which is the intersection of the hyperplane orthogonal to the vector $b_4 \omega_4$ and the sphere $S^3$ defined by the orbit $W(H_4)\Lambda$. Then the branching of an arbitrary orbit $W(H_4)\Lambda$ can be given as

$$(a_1,a_2,a_3,a_4) = (a_1,a_2,a_3)\{\pm\frac{1}{2}(\tau^3 a_1 + 2\tau^2 a_2 + (\tau+2)a_3 + 2a_4)\}$$

$$+(\tau a_1 + a_2 + a_3 + a_4, a_2 + a_3, \tau a_1 + \tau a_2)\{\pm\frac{1}{2}(a_1 + \tau a_2 + a_3 + a_4)\} \quad (20)$$

$$+\cdots(116 \text{ more vectors})$$

From now on we simply use the vector notation $\Lambda$ rather than $W(G)\Lambda$ for the designation of the orbit. The decomposition in (20) is obtained by a computer calculation. We will only list the decomposition of those polytopes where $a_1, a_2, a_3, a_4$ take certain integer values. Before we proceed further we wish to classify the solids possessing the icosahedral symmetry $W(H_3)$. The orbits obtained from the vectors $(1,0,0)$, $(0,1,0)$, $(0,0,1)$ represent respectively dodecahedron, icosidodecahedron and icosahedron [20]. The orbits obtained from the vectors $(b_1, b_2, 0)$ and $(0, b_2, b_3)$ represent respectively a quasi regular truncated dodecahedron with edges of lengths $b_1$ and $b_2$ and a quasi regular truncated icosahedron with edge lengths $b_2$ and $b_3$. If $b_1 = b_2$ in the first vector and $b_2 = b_3$ in the second vector then their orbits represent respectively the regular truncated dodecahedron and the regular truncated icosahedron. Finally the orbits of the vectors $(b_1, 0, b_3)$ and $(b_1, b_2, b_3)$ respectively represent a quasi regular small rhombicosidodecahedron with edge lengths $b_1$ and $b_3$ and a quasi regular great rhombicosidodecahedron with three edge lengths $b_1, b_2$ and $b_3$. When the components of the vectors are equal we obtain the Archimedean solids including the small and great rhombicosidodecahedra.

Now we display the decompositions of some of the 4D solids under the $W(H_3)$ symmetry.

*3.1. The 600-cell*

The orbit $W(H_4)(0,0,0,1)$ represents the 600-cell with 120 vertices. Each cell is a tetrahedron, 20 of which are meeting at one vertex. The orbit can be decomposed under the subgroup $W(H_3)$ as follows:

$$(0,0,0,1) = (0,0,0)(\pm 1) + (0,0,1)(\pm\frac{\tau}{2}) + (0,0,\tau)(\pm\frac{\sigma}{2}) + (1,0,0)(\pm\frac{1}{2}) + (0,1,0)(0). \quad (21)$$

From now on we assume that the orbit is represented by a vector where the components are taken to be positive numbers.
Actually the orbit $W(H_4)(0,0,0,1)$ can be represented by the set of quaternions $I$ when the vector $\omega_4 = -\sqrt{2}\tau$ is multiplied by $\dfrac{\sigma}{\sqrt{2}}$, then, in turn, the orbits in (21) represent the conjugacy classes of the group $I$ given in Table 1. The orbits $(0,0,0)(\pm 1)$ represent two poles of the sphere $S^3$, the other orbits respectively represent icosahedra, dodecahedra and an icosidodecahedron.

*3.2. The 120-cell*

It is the dual of the 600-cell and its cells are the dodecahedra, 4 of which are meeting at one point. It is represented by the orbit $W(H_4)(1,0,0,0)$ which we have studied elsewhere in the context of quaternions [17] and [21]. It is decomposed as follows:

$$(1,0,0,0) = (1,0,0)(\pm\frac{\tau^3}{2}) + (\tau,0,0)(\pm\frac{\tau+2}{2}) + (\tau^2,0,0)(\pm\frac{\sigma}{2}) + (0,\tau,0)(\pm\tau) + (0,1,\tau)(\pm\frac{\tau^2}{2})$$
$$+ (0,\tau,1)(\pm\frac{\tau}{2}) + (\tau,0,\tau)(\pm\frac{1}{2}) + (1,0,\tau^2)(0).$$
(22)

All vectors can be converted to unit quaternions by multiplying each vector with $\frac{\sigma}{\sqrt{2}}$ then the 600 vertices are represented by the set of quaternions $J = \sum_{i,j=0}^{4} \oplus p^i \, T' \, p^j$ where $p^5 = 1$.

*3.3. The 720- cell*

It is described by the orbit $W(H_4)(0,1,0,0)$ which has 1,200 vertices. The polytope consists of two types of cells: tetrahedra (600) and icosidodecahedra (120). Its decomposition under the group $W(H_3)$ is given by

$$(0,1,0,0) = (\tau,0,0)(\pm\frac{3\tau}{2}) + (0,1,0)(\pm\tau^2) + (0,\tau^2,0)(\pm 1) + (1,10)(\pm\frac{\tau^3}{2}) + (0,\tau,1)(\pm\frac{\tau+2}{2})$$
$$+ (0,\tau,\tau)(\pm\frac{\tau^2}{2}) + (1,0,\tau^2)(\pm\tau) + (\tau^2,0,1)(\pm\frac{1}{2}) + (\tau,0,\tau^2)(\pm\frac{\sigma}{2}) + (1,1,\tau)(\pm\frac{\tau}{2})$$
$$+ (2\tau,0,0)(0) + (0,1,2\tau)(0).$$
(23)

*3.4. The $720'-cell$*

It is described by the orbit $W(H_4)(0,0,1,0)$ which has 720 vertices. It consists of 600 octahedra and 120 icosahedra. It is the unique 4D Archimedean polytope with $W(H_4)$ symmetry in the sense that its cells are made of Platonic solids. It decomposes under the group $W(H_3)$ as follows

$$(0,0,1,0) = (0,0,1)(\pm\frac{\tau+2}{2}) + (0,0,\tau^2)(\pm\frac{\tau-\sigma}{2}) + (0,1,0)(\pm\tau) + (0,0,2\tau)(0)$$
$$+ (0,1,\tau)(\pm\frac{\sigma}{2}) + (1,0,1)(\pm\frac{\tau^2}{2}) + (1,0,\tau)(\pm\frac{\tau}{2}) + (1,1,0)(\pm\frac{1}{2})$$
$$+ (\tau,0,1)(0) + (0,\tau,0)(\pm 1).$$
(24)

One can easily interpret the type of projected polyhedra by consulting the above discussions.

## 4. The $W(H_4)$ polytopes and branching under the group $W(A_4)$

The right-coset decomposition of the group $W(H_4)$ under the subgroup $W(A_4)$ can be given by

$$W(H_4) = \sum_{i=1}^{120} \oplus W(A_4) g^i \tag{25}$$

where the right coset elements form the same subgroup $g^i \in [I,1]$. The vectors to which $W(A_4)$ will act can be taken as $\Lambda^{(i)} \equiv g^i \Lambda$, $i = 1,2,...,120$.

The group $W(A_4)$ acts on an arbitrary vector $(b_1,b_2,b_3,b_4) \equiv b_1 v_1 + b_2 v_2 + b_3 v_3 + b_4 v_4$ where $(v_i, \alpha_j) = \delta_{ij}$ with $i,j = 1,2,3,4$ and the Cartan matrix $(C)_{ij} = (\alpha_i, \alpha_j)$ and its inverse $(C^{-1})_{ij} = (v_i, v_j)$ of the group $W(A_4)$ are given by

$$C_{A_4} = \begin{pmatrix} 2 & -1 & 0 & 0 \\ -1 & 2 & -1 & 0 \\ 0 & -1 & 2 & -1 \\ 0 & 0 & -1 & 2 \end{pmatrix}, \quad (C^{-1})_{A_4} = \frac{1}{5}\begin{pmatrix} 4 & 3 & 2 & 1 \\ 3 & 6 & 4 & 2 \\ 2 & 4 & 6 & 3 \\ 1 & 2 & 3 & 4 \end{pmatrix}. \tag{26}$$

The dual vectors $v_i$ can be written in terms of quaternions as follows

$$v_1 = \frac{1}{\sqrt{10}}(-\sqrt{5} + \tau e_2 - \sigma e_3), \quad v_2 = \frac{1}{\sqrt{10}}(2\tau e_2 - 2\sigma e_3),$$

$$v_3 = \frac{1}{\sqrt{10}}(-\sqrt{5} e_1 + \tau^2 e_2 - \sigma^2 e_3), \quad v_4 = \frac{1}{\sqrt{10}}(2e_2 - 2e_3) \equiv -\frac{2}{\sqrt{5}} c. \tag{27}$$

We have to express the $W(A_4)$ orbits in the basis of the vectors $v_i$. It is easy to obtain the relation $\omega_i = D_{ij} v_j$ where the matrix $D$ is given by

$$D = \begin{pmatrix} \tau^4 & -2\tau^2 & 1 & \tau \\ 2\tau^3 & -(3\tau+1) & 0 & \tau^2 \\ (3\tau+1) & -\tau^3 & 0 & 1 \\ 2\tau & -\tau & 0 & 0 \end{pmatrix}. \tag{28}$$

Any vector $\Lambda^{(i)} \equiv g^i \Lambda$ is now in the form $(b_1, b_2, b_3, b_4) \equiv b_1 v_1 + b_2 v_2 + b_3 v_3 + b_4 v_4$.

The components of these vectors do not in general satisfy the relation $b_1 \geq 0, b_2 \geq 0, b_3 \geq 0, b_4 \geq 0$. Nevertheless there is always a vector in the orbit $W(A_4)\Lambda^{(i)}$ where the relations $b_1 \geq 0, b_2 \geq 0, b_3 \geq 0, b_4 \geq 0$ are satisfied. Then the branching of an arbitrary orbit $W(H_4)\Lambda$ can be given as

$$(a_1,a_2,a_3,a_4) = \{a_2v_1 + a_3v_2 + a_4v_3 +[(3\tau+1)a_1 + \tau^4 a_2 + \tau^3 a_3 + \tau a_4]v_4\} + \{v_1 \leftrightarrow v_4; v_2 \leftrightarrow v_3\} + \cdots.$$
$$+(118 \text{ more vectors like this decomposition}).$$
(29)

In the above decomposition all the vectors come in pairs as shown in (29). Now we display the decompositions of some of those 4D polytopes with $W(H_4)$ symmetry under the subgroup $W(A_4)$.

*4.1. The 600-cell*
$$(0,0,0,1) = (0,0,1,\tau) + (\tau,1,0,0) + (\tau,0,0,\tau) + (1,0,\tau,0) + (0,\tau,0,1). \quad (30)$$
The first three are the polytopes with 20 vertices each and the last two polytopes have 30 vertices each. In particular the vertices of the polytope $\tau(1,0,0,1)$ represent a scaled non-zero root system of the Coxeter–Weyl group $W(A_4)$.

*4.2. The 120-cell*
$$(1,0,0,0) = (3\tau+1,0,0,0) + (0,0,0,3\tau+1) + (1,\tau^3,0,0) + (0,0,\tau^3,1) + (\tau,0,0,2\tau^2) + (2\tau^2,0,0,\tau)$$
$$+ (0,\tau^2,\tau^2,0) + (1,\tau,0,\tau^3) + (\tau^3,0,\tau,1) + (\tau,\tau^2,\tau,0) + (0,\tau,\tau^2,\tau) + (\tau^2,\tau,1,\tau) + (\tau,1,\tau,\tau^2).$$
(31)

*4.3. The 720-cell*
$$(0,1,0,0) = (0,3\tau+1,0,0) + (0,0,3\tau+1,0) + (\tau^4,0,0,1) + (1,0,0,\tau^4) + (2\tau^2,0,\tau^2,0) + (0,\tau^2,0,2\tau^2)$$
$$+(\tau,\tau^3,1,0) + (0,1,\tau^3,\tau) + (1,2\tau,\tau^2,0) + (0,\tau^2,2\tau,1) + (\tau^2,0,\tau^2,2\tau) + (2\tau,\tau^2,0,\tau^2) + (\tau,1,0,3\tau+1)$$
$$+(3\tau+1,0,1,\tau) + (1,\tau,\tau^2,\tau^2) + (\tau^2,\tau^2,\tau,1) + (\tau,\tau,1,\tau^3) + (\tau^3,1,\tau,\tau) + (\tau^2,\tau,\tau,\tau^2).$$
(32)

*4.4. The $720'-cell$*
$$(0,01,0) = (\tau^3,0,1,0) + (0,1,0,\tau^3) + (0,2\tau,1,0) + (0,1,2\tau,0) + (\tau,\tau^2,0,1) + (1,0,\tau^2,\tau) + (\tau^2,0,\tau,\tau)$$
$$+(\tau,\tau,0,\tau^2) + (\tau^2,1,\tau,0) + (0,\tau,1,\tau^2) + (2\tau,1,0,\tau) + (\tau,0,1,2\tau) + (1,\tau,\tau,1).$$
(33)

If we wish to decompose the $W(H_4)$ orbits under the subgroup $W(A_3)$ then we use the decomposition of the $W(A_4)$ orbits under the group $W(A_3)$ as shown below [22]

$$O(a_1,a_2,a_3,a_4) = O(a_1,a_2,a_3)(-(a_1+2a_2+3a_3+4a_4)) + O(a_1,a_2,(a_3+a_4))(-(a_1+2a_2+3a_3-a_4))$$
$$+ O(a_1,(a_2+a_3),a_4)(-(a_1+2a_2-2a_3-a_4)) + O((a_1+a_2),a_3,a_4)(-(a_1-3a_2-2a_3-a_4))$$
$$+ O(a_2,a_3,a_4)(4a_1+3a_2+2a_3+a_4).$$
(34)

For example, the 600-cell with the 120 vertices decomposes under the group $W(A_3)$ as

$$(0,0,0,1) = (1,0,0)(4\tau+3) + (0,0,1)(-(4\tau+3)) + (\tau,0,0)(-5\tau) + (0,0,\tau)(5\tau) + +(\tau^2,0,0)(2+\sigma)$$
$$+(0,0,\tau^2)(-(2+\sigma)) + (0,\tau,0)(\pm(2\tau+4)) + (1,0,\tau)(-(3\tau+1)) + (\tau,0,1)(3\tau+1)$$
$$+(0,\tau,1)(-\sqrt{5}) + (1,\tau,0)(\sqrt{5}) + (\tau,1,0)(-(\tau+2)) + (0,1,\tau)(\tau+2) + (\tau,0,\tau)(0).$$
(35)

## 5. Dual polytopes of the uniform polytopes of the Coxeter-Weyl group $W(H_4)$

In the references [20] and [23] we have proved that all the Catalan solids can be derived from the Coxeter diagrams $A_3, B_3, H_3$ with a simple technique which we will also apply for the constructions of the dual polytopes of the 4D polytopes. To obtain the vertices of the dual polytope of a given polytope one determines the centers of the cells joined to the vertex $(a_1, a_2, a_3, a_4)$. The relative magnitudes of these vectors are determined from the fact that the hyperplane formed by the vectors representing the centers of the cells are orthogonal to the vertex $(a_1, a_2, a_3, a_4)$. Let us recall that the dual polytopes are cell transitive similar to the Catalan solids which are face transitive. In what follows we study the duals of the $W(H_4)$ uniform polytopes in turn.

### 5.1. Dual polytope of the 600-cell $O(0001) = O(\omega_4)$

The tetrahedral subgroup $W(A_3)$ generated by $W(A_3) = \langle r_2, r_3, r_4 \rangle$ acting on the vector $(0,0,0,1)$ generates a tetrahedron whose center can be represented by the vector $\omega_1$ as the group $W(A_3) = \langle r_2, r_3, r_4 \rangle$ leaves the vector $\omega_1$ invariant. The number of tetrahedra joining to the vertex $\omega_4$ is given by the formula

$$\frac{|W(H_4)|}{|W(A_3)|} \times \frac{\text{number of vertices of the tedrahedron}}{\text{number of vertices of the polytope}} = \frac{120 \times 120}{24} \times \frac{4}{120} = 20.$$
(36)

This formula has an interesting group theoretical interpretation. The orbit which includes the vector $\omega_1 = (1,0,0,0)$ can be generated by the icosahedral group $W(H_3) = \langle r_1, r_2, r_3 \rangle$ and represents a dodecahedron. Note that the dihedral subgroup $D_3 = \langle r_2, r_3 \rangle$ leaves the vector $\omega_1$ invariant. A left coset decomposition $W(H_3) = \sum_{i=1}^{20} \oplus g^i D_3$ will generate the vertices of the dodecahedron where $g^i \in W(H_3)/D_3$. At each vertex of the 600-cell there exist 20 tetrahedra, each of which, is generated by a conjugate group $g^i W(A_3)(g^i)^{-1}$, $i = 1, 2, ..., 20$.
Similarly the left coset decomposition of the tetrahedral group can be written as $W(A_3) = \sum_{i=1}^{4} f^i D_3$ where the group $D_3 = \langle r_2, r_3 \rangle$ leaves the vectors $\omega_1 = (1,0,0,0)$ and $\omega_4 = (0,0,0,1)$ invariant. It is those conjugate groups $f^i W(H_3)(f^i)^{-1}$ ($i = 1, 2, 3, 4$) generate four dodecahedra when applied on the vector $\omega_1$ whose centers form four tetrahedra. The number of vertices of the polytope $W(H_4)\omega_1$ is 600 which are equal to the number of left cossets (index) of the group $W(A_3)$ in the group $W(H_4)$. The number of dodecahedral cells of the polytope $W(H_4)\omega_1$ is 120 since it is equal to the number of conjugate groups of the group $W(H_3)$ in the group $W(H_4)$. Therefore the 120-cell is the dual of the 600-cell and the vice versa.

*5.2. Dual polytope of the 720-cell $O(0,1,0,0) = O(\omega_2)$*

This uniform polytope consists of the cells tetrahedra and the icosidodecahedra since the subgroup $W(A_3) = \langle r_2, r_3, r_4 \rangle$ acting on the vector $(0,1,0,0)$ generates a tetrahedron and the subgroup $W(H_3)\omega_2 = W(H_3)(0,1,0,0)$ generates an icosidodecahedron with 30 vertices for the index of the group $C_2 \times C_2 = \langle r_1, r_3 \rangle$ leaving the vector $\omega_2$ invariant is equal to $\frac{|W(H_3)|}{|C_2 \times C_2|} = 30$. Actually, the vector $\omega_2$ is left invariant by the group of generators $D_3 \times C_2 = \langle r_1, r_3, r_4 \rangle$ defining the symmetry of the triangular prism of order 12. Therefore the number of the vertices of the polytope is equal to the index $\frac{|W(H_4)|}{|D_3 \times C_2|} = 1200$. Total number of cells are given by the sum of indices

$$\frac{|W(H_4)|}{|W(A_3)|} + \frac{|W(H_4)|}{|W(H_3)|} = 600 + 120.$$
(37)

The number of cells which share the vector $\omega_2$ as a vertex is determined by the formula similar to the one in equation (36) which leads to 2 tetrahedra and 3 icosidodecahedra at the vertex $\omega_2$. Then the centers of these polyhedra are given by the vectors up to a relative scale factor $\lambda$:

The centers of the tetrahedra : $\lambda\omega_1, \lambda r_1\omega_1$

The centers of the icosidodecahedra : $\omega_4, r_3r_4\omega_4, r_4r_3\omega_4$. (38)

These vertices define a hyperplane orthogonal to the vector $\omega_2$. Therefore one can determine the scale factor $\lambda$ from the equation $(\lambda\omega_1 - \omega_4).\omega_2 = 0$ leading to the value $\lambda = \frac{2}{3\tau}$. The set of vertices of the dual polytope is the union of the orbits $\frac{2}{3\tau}O(1,0,0,0) \oplus O(0,0,0,1)$. They define two concentric $S^3$ spheres with the fraction of radii $R_4/R_1 \approx 1.061$ where $R_1$ and $R_4$ represent the radii of the spheres determined by the set of vectors in the orbits respectively.

The last three vectors in (38) define an equilateral triangle whose vertices are permuted by the generator $r_3r_4$ with $(r_3r_4)^3 = 1$. The vector $\lambda(\omega_1 - r_1\omega_1) = \lambda\alpha_1$ is orthogonal to the plane of the triangle. Therefore the solid is a dipyramid with triangular base with the edge lengths $\sqrt{2} \approx 1.41$ and $|\lambda\omega_1 - \omega_4| \approx 0.867$. Let us introduce new set of quaternionic unit vectors $p_0 = \frac{\omega_2}{|\omega_2|}$, $p_1 = e_1 p_0$, $p_2 = e_2 p_0$, $p_3 = e_3 p_0$. Since the dipyramid lies in the hyperplane orthogonal to the vector $p_0 = \frac{\omega_2}{|\omega_2|}$ the five vectors in (38), when expressed in terms of quaternions in the basis $p_0$, $p_1$, $p_2$, $p_3$, will have the same $p_0$ components. If we disregard the $p_0$ components and an overall scale factor the vertices of the dipyramid are given by the set of vectors in 3-dimensional space as follows

$$(0,0,1), \quad \frac{1}{2}(\sigma, \tau, -1), \quad \frac{1}{2}(-\sigma, -\tau, -1), \quad \frac{1}{3}(-1, -\sigma^2, 0), \quad \frac{1}{3}(1, \sigma^2, 0). \tag{39}$$

Plots of this dipyramid are shown in Figure 3.

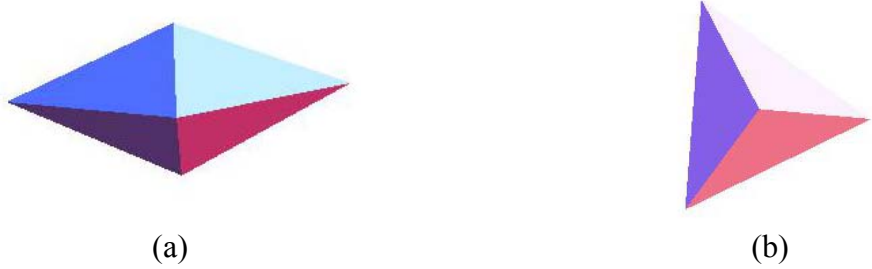

(a)          (b)

**Figure 3.** (a) The dipyramid, a typical cell of the polytope $\frac{2}{3\tau} O(1,0,0,0) \oplus O(0,0,0,1)$.
(b) Top view.

## 5.3. Dual polytope of the $720'$-cell $O(0,0,1,0) = O(\omega_3)$

This polytope has 720 vertices equal to the index of the group $D_5 \times C_2$ of order 20 in the group $W(H_4)$. The cells of this polytope consist of two types of Platonic solids: icosahedral and octahedral cells.

The subgroup $W(H_3) = \langle r_1, r_2, r_3 \rangle$ acting on the vector $\omega_3$ will generate the vertices of an icosahedron. Similarly the subgroup $W(A_3) = \langle r_2, r_3, r_4 \rangle$ acting on the vector $\omega_3$ describes an octahedron. Total number of cells is 720 with 120 icosahedral cells and 600 octahedral cells. The number of cells sharing the vector $\omega_3$ as a vertex consists of 2 icosahedra and 5 octahedra. The centers of the octahedral cells can be taken to be the vectors $\omega_1$, $r_1 r_2 \omega_1$, $(r_1 r_2)^2 \omega_1$, $(r_1 r_2)^3 \omega_1$, $(r_1 r_2)^4 \omega_1$ and the centers of the icosahedral cells are determined to be the vectors $\omega_4$ and $r_4 \omega_4$ up to a scale factor. Similar to the argument raised in Section 5.2 the scale factor of the last two vectors are determined from the relation $(\lambda\omega_4 - \omega_1).\omega_3 = 0$ with $\lambda = \dfrac{2\tau}{2+\sigma}$. These seven vectors describe a dipyramid with pentagonal base with edge length $\sqrt{2} \approx 1.41$ and the other edges of length $|\lambda\omega_4 - \omega_1| = \sqrt{\dfrac{8}{5}} \approx 1.26$. The dual polytope consist of 720 vertices represented by the union of two orbits

$$O(1,0,0,0) \oplus \dfrac{2\tau}{2+\sigma} O(0,0,0,1). \tag{40}$$

These two orbits define two concentric spheres $S^3$ with the ratio of the radii $R_4/R_1 \approx 1.023$. Expressing the vertices in the basis of unit quaternions $p_0 = \dfrac{\omega_3}{|\omega_3|}$, $p_1 = e_1 p_0$, $p_2 = e_2 p_0$, $p_3 = e_3 p_0$ and then deleting the common components of the vector $p_0 = \dfrac{\omega_3}{|\omega_3|}$, the vertices of the dipyramid are given by the set of vectors

$$(0,1,\tau),\ (0,-1,-\tau),\ \dfrac{\sqrt{5}}{2}(-1,-\sigma,-\sigma^2),\ \dfrac{\sqrt{5}}{2}(\sigma^2,1,\sigma),\ \dfrac{\sqrt{5}}{2}(-2\sigma,0,0),$$
$$\dfrac{\sqrt{5}}{2}(\sigma^2,-1,-\sigma),\ \dfrac{\sqrt{5}}{2}(-1,\sigma,\sigma^2). \tag{41}$$

Plots of these solid are shown in Figure 4.

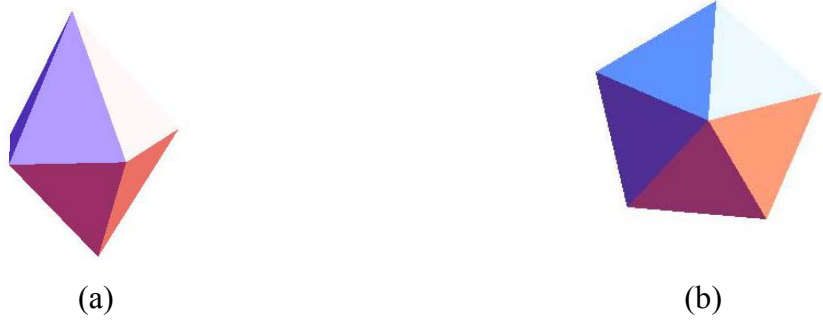

(a)          (b)

**Figure 4.** (a) The dipyramid, a typical cell of the polytope $O(1,0,0,0) \oplus \dfrac{2\tau}{3+\sigma} O(0,0,0,1)$.
(b) Top view.

This polytope is the projection of the Voronoi cell of $W(E_8)$ [24] with 19,440 vertices into 4D space with the residual symmetry $W(H_4)$.

*5.4. Dual polytope of the polytope $O(1,0,0,1) = O(\omega_1 + \omega_4)$*

The generators $r_2$ and $r_3$ fixing the vector $\Lambda = \omega_1 + \omega_4$ form a dihedral group of order 6 indicating that the polytope has 2400 vertices. There are four types of cells sharing the same vertex. The subgroup $W(H_3) = \langle r_1, r_2, r_3 \rangle$ acting on the vector $\omega_1 + \omega_4$ generates a dodecahedron as it leaves the vector $\omega_4$ invariant. The number of dodecahedral cells is then 120. The subgroup $W(A_3) = \langle r_2, r_3, r_4 \rangle$ generates a tetrahedron with a total of 600. The subgroup $D_5 \times C_2 = \langle r_1, r_2, r_4 \rangle$ generates a pentagonal prism. There are 720 cells of this type. Finally the subgroup $D_3 \times C_2 = \langle r_1, r_3, r_4 \rangle$ generates a cell with a total of 1200 triangular prisms. Therefore the total number of cells is 2640.

The number of cells sharing the same vertex and the vectors representing their centers are determined using a formula similar to (36) as

Dodecahedral cell : 1 and its center represented by $\omega_4$
Tetrahedral cell  : 1 and its center represented by $\omega_1$
Triangular prisms : 3 and their centers represented by $\omega_2$, $(r_2 r_3)\omega_2$, $(r_2 r_3)^2 \omega_2$     (42)
Pentagonal prisms : 3 and their centers represented by $\omega_3$, $(r_2 r_3)\omega_3$, $(r_2 r_3)^2 \omega_3$

These vectors are defined up to some scale vectors. Now if we let $\lambda \omega_3, (r_2 r_3)\lambda \omega_3, (r_2 r_3)^2 \lambda \omega_3, \rho \omega_1$ and $\eta \omega_4$, then the scale factors are determined from the equations

$(\lambda \omega_3 - \omega_2).\Lambda = 0$, $(\rho \omega_1 - \omega_2).\Lambda = 0$ and $(\eta \omega_4 - \omega_2).\Lambda = 0$ as $\lambda = \dfrac{\tau^4}{\tau+3}, \rho = \dfrac{\tau^4}{\tau+4}$ and $\eta = \dfrac{\tau^4}{\sigma+3}$.

Above solutions are unique to determine the hyperplane orthogonal to the vector $\Lambda = \omega_1 + \omega_4$.

The 2640 vertices of the dual polytope is defined as the union of the orbits

$$\rho\, O(1,0,0,0) \oplus O(0,1,0,0) \oplus \lambda\, O(0,0,1,0) \oplus \eta\, O(0,0,0,1). \quad (43)$$

These vertices lie on the concentric four $S^3$ spheres with the ratio of the radii

$$R_2 : R_3 : R_4 : R_1 \approx 2.45 : 2.47 : 2.52 : 2.97. \quad (44)$$

The dual polytope has 1200 cells each of which is a solid with 8 vertices. These 8 vertices can be expressed in the basis of unit quaternions defined by $p_0 = \dfrac{\omega_1 + \omega_4}{|\omega_1 + \omega_4|}$, $p_1 = e_1 p_0$, $p_2 = e_2 p_0$, $p_3 = e_3 p_0$. Excluding the common components in the direction of $p_0$ they can be determined in 3 dimensions as follows:

$\rho\omega_1 = \rho[\sigma^2,0,1]$, $\quad \omega_2 = [-1,-\sigma^2,-2\sigma]$, $\quad r_2r_3\omega_2 = [-\sigma,\sigma+2,-\sigma]$, $(r_2r_3)^2\omega_2 = [-2\sigma,-1,\sigma^2]$,
$\lambda\omega_3 = \lambda[-1,1,-\sigma^3]$, $\lambda(r_3r_2)\omega_3 = \lambda[-\sigma^2,-(\sigma+2),0]$, $\lambda r_2r_3\omega_3 = \lambda[-2\sigma,\sigma^2,\sigma]$,
$\eta\omega_4 = \eta[-\sigma^2,0,-1]$.
(45)

The solid defined by these 8 vertices has a dihedral symmetry $D_3$. Six faces consist of two types of kites, big and small. Three small kites meet at the top vertex and three big kites join to the bottom vertex. At the other six vertices either two small kites and one big kite or two big kites and one small kite get together. Plots of this solid are shown in Figure 5.

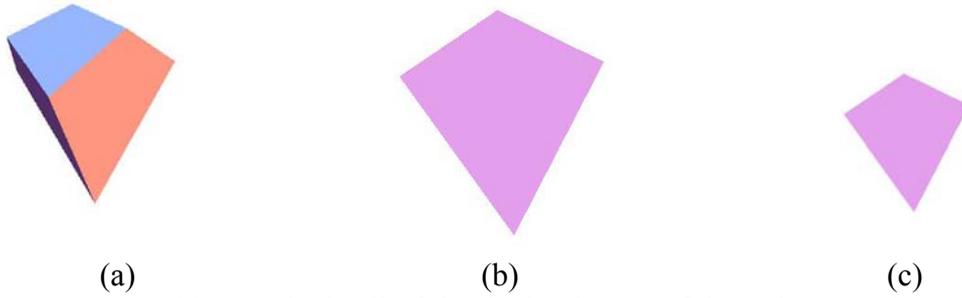

(a)                      (b)                     (c)

**Figure 5.** (a) A typical cell of the dual polytope of the polytope $O(1,0,0,1)$.
(b) Big kite view. (c) Small kite view.

*5.5. Dual polytope of the polytope $O(0,1,0,1) = O(\omega_2 + \omega_4)$*

This polytope has 3600 vertices and 1440 cells. The cells consist of 120 icosidodecahedra, 600 cuboctahedra and 720 pentagonal prisms.

The number of cells sharing the same vertex $(0,1,0,1) = \omega_2 + \omega_4$ and the vectors representing their centers are determined as follows:

Icosidodecahedron : 1 and its center represented by $\lambda\omega_4$
Cuboctahedra : 2 and their centers represented by $\rho\omega_1, \rho r_1\omega_1$
Pentagonal prisms : 2 and their centers represented by $\omega_3, r_3\omega_3$.

The scale vectors can be determined as $\lambda = \dfrac{4\tau+13}{10}, \rho = \dfrac{7\tau-8}{4}$.

The vertices of the dual polytope can be represented by the union of the following orbits

$$\rho\, O(1,0,0,0) \oplus O(0,0,1,0) \oplus \lambda O(0,0,0,1). \tag{46}$$

These vertices lie on the concentric three $S^3$ spheres with the ratio of the radii
$$R_3 : R_1 : R_4 \approx 1.6625 : 1.6631 : 1.7019. \tag{47}$$

The 5 vertices can be expressed in the basis of unit quaternions defined by $p_0 = \dfrac{\omega_2 + \omega_4}{|\omega_2 + \omega_4|}$, $p_1 = e_1 p_0$, $p_2 = e_2 p_0$, $p_3 = e_3 p_0$. Excluding the common components in the direction of $p_0$ they can be determined as follows:

$$\rho\omega_1 = \rho[-\sqrt{5}, \sigma, -\tau],\ \rho r_1\omega_1 = \rho[\sqrt{5}, -\sigma, -\tau],\ \omega_3 = [-\sigma, -\sqrt{5}, 0],\ r_3\omega_3 = [\sigma, \sqrt{5}, 0],\ \lambda\omega_4 = \lambda[0,0,2]. \quad (48)$$

This cell is a dipyramid with a base of isosceles triangle possessing the symmetry $C_2 \times C_2 = \langle r_1, r_3 \rangle$. Plots of this solid are shown in Figure 6.

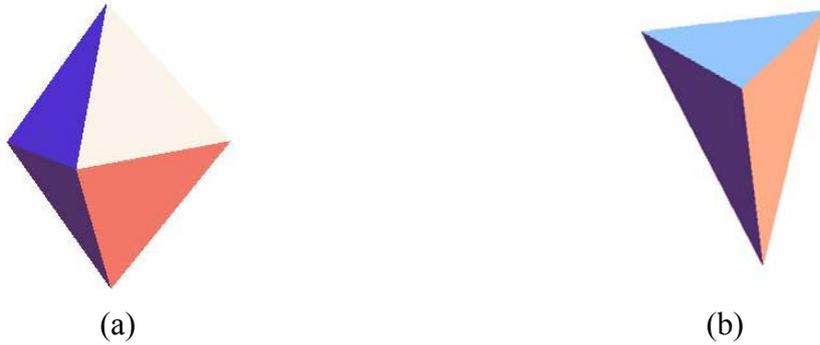

(a) (b)

**Figure 6.** (a) A typical cell of the dual polytope of the polytope $O(0,1,0,1)$. (b) Top view.

*5.6. Dual polytope of the polytope $O(0,0,1,1) = O(\omega_3 + \omega_4)$*

This polytope has 1440 vertices and 720 cells. The cells consist of 120 icosahedra and 600 truncated tetrahedra. The number of cells sharing the same vertex $(0,0,1,1) = \omega_3 + \omega_4$ and the vectors representing their centers are determined as follows:

Icosahedra : 1 and its center represented by $\lambda\omega_4$
Truncated tetrahedra : 5 and centers represented by $\omega_1, r_1 r_2\omega_1, (r_1 r_2)^2\omega_1, (r_1 r_2)^3\omega_1, (r_1 r_2)^4\omega_1$.

The scale vector can be determined as $\lambda = \dfrac{21\tau + 9}{19}$.

This is a pyramid with a pentagonal base and edge lengths 1.41 and 2.01. The vertices of the dual polytope can be represented as the union of the following orbits

$$O(1,0,0,0) \oplus \lambda O(0,0,0,1). \quad (49)$$

These orbits define two concentric $S^3$ spheres with the ratio of the radii

$$R_1 : R_4 \approx 1.012. \quad (50)$$

The 6 vertices can be expressed in the basis of unit quaternions defined by $p_0 = \dfrac{\omega_3 + \omega_4}{|\omega_3 + \omega_4|}$, $p_1 = e_1 p_0$, $p_2 = e_2 p_0$, $p_3 = e_3 p_0$. Excluding the common components in the direction of $p_0$ they can be determined as follows:

$$\omega_1 = [\sqrt{5}, -1, \sqrt{5}], \ r_1 r_2 \omega_1 = [2\sigma, -\tau, \tau^2], (r_1 r_2)^2 \omega_1 = [-3, 1, 1], (r_1 r_2)^3 \omega_1 = [\sigma, 2\tau, -\sigma^2],$$
$$(r_1 r_2)^4 \omega_1 = [\tau^2, 2, \sigma^2], \lambda \omega_4 = \lambda[0, \sigma, -1]. \tag{51}$$

It is a pyramid with a regular pentagonal base as shown in Figure 7.

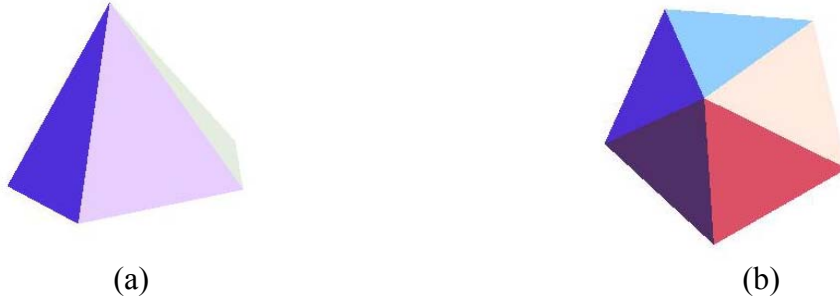

(a)                          (b)

**Figure 7**. (a) A typical cell of the dual polytope of the polytope $O(0,0,1,1)$. (b) Top view.

*5.7. Dual polytope of the polytope $O(0,1,1,0) = O(\omega_2 + \omega_3)$*

This polytope has 3600 vertices and 720 cells. The cells consist of 120 truncated icosahedra and 600 truncated tetrahedra. The number of cells sharing the same vertex $(0,1,1,0) = \omega_2 + \omega_3$ and the vectors representing their centers are determined as follows:

Truncated icosahedra     : 2 and their centers represented by $\lambda \omega_4, \lambda r_4 \omega_4$
Truncated tetrahedra     : 2 and their centers represented by $\omega_1, r_1 \omega_1$.

The scale factor can be determined as $\lambda = \dfrac{5\tau}{4+\sigma}$. This is a solid with four vertices and with two types of faces represented by two inequivalent isosceles triangles possessing a symmetry $C_2 \times C_2 = \langle r_1, r_4 \rangle$ of order 4. The vertices of the dual polytope can be represented as the union of the following orbits

$$O(1,0,0,0) \oplus \lambda \ O(0,0,0,1). \tag{52}$$

These orbits define two concentric $S^3$ spheres with the ratio of the radii
$$R_1 : R_4 \approx 0.957. \tag{53}$$

The 4 vertices can be expressed in the basis of unit quaternions defined by $p_0 = \dfrac{\omega_2 + \omega_3}{|\omega_2 + \omega_3|}$, $p_1 = e_1 p_0$, $p_2 = e_2 p_0$, $p_3 = e_3 p_0$. Excluding the common components in the direction of $p_0$ they can be determined as follows:

$$\omega_1 = [-2\tau^2, -\sigma, -1], \ r_1 \omega_1 = [\tau+2, \tau+2, -\tau], \lambda \omega_4 = \lambda[0,1,3\tau], \lambda r_4 \omega_4 = \lambda[1, -(\tau+2), -2\tau]. \tag{54}$$

A plot of this solid is shown in Figure 8.

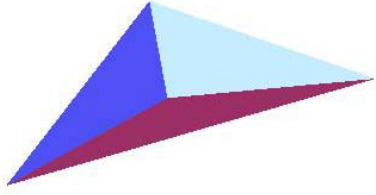

**Figure 8.** A typical cell of the dual polytope of the polytope $O(0,1,1,0)$.

*5.8. Dual polytope of the polytope* $O(1,0,1,0) = O(\omega_1 + \omega_3)$

This polytope has 3600 vertices and 1920 cells. The cells consist of 120 small rhombicosidodecahedra, 600 octahedra and 1200 triangular prisms. The number of cells sharing the same vertex $(1,0,1,0) = \omega_1 + \omega_3$ and the vectors representing their centers are determined as follows:

Small rhombicosidodecahedra : 2 and centers represented by $\omega_4, r_4\omega_4$

Octahedron : 1 and its center represented by $\lambda\omega_1$

Triangular prisms : 2 and centers represented by $\rho\omega_2, \rho r_2\omega_2$.

The scale factors can be determined as $\lambda = \dfrac{3}{4+2\tau}$ and $\rho = \dfrac{3}{4+3\tau}$. This is a dipyramid with an isosceles triangular base left invariant under the symmetry $C_2 \times C_2 = \langle r_2, r_4 \rangle$ of order 4. The vertices of the dual polytope can be represented as the union of the following orbits

$$\lambda\, O(1,0,0,0) \oplus \rho\, O(0,1,0,0) \oplus O(0,0,0,1). \tag{55}$$

These orbits define three concentric $S^3$ spheres with the ratio of the radii

$$R_1 : R_2 : R_4 \approx 0.9991 : 0.9495 : 1.0000. \tag{56}$$

The 5 vertices can be expressed in the basis of unit quaternions defined by $p_0 = \dfrac{\omega_1 + \omega_3}{|\omega_1 + \omega_3|}$, $p_1 = e_1 p_0$, $p_2 = e_2 p_0$, $p_3 = e_3 p_0$. Excluding the common component in the direction of $p_0$ they can be determined as follows:

$$\lambda\omega_1 = \lambda[-\tau^2, \tau, -1],\ \rho\omega_2 = \rho[\tau, \tau+2, -\tau],\ \rho r_2\omega_2 = \rho[-2\tau, -\tau^2, 1],\ \omega_4 = [1, 1, \tau^3],$$
$$r_4\omega_4 = [\tau, -\tau^2, -2\tau]. \tag{57}$$

Plots of this solid are shown in Figure 9.

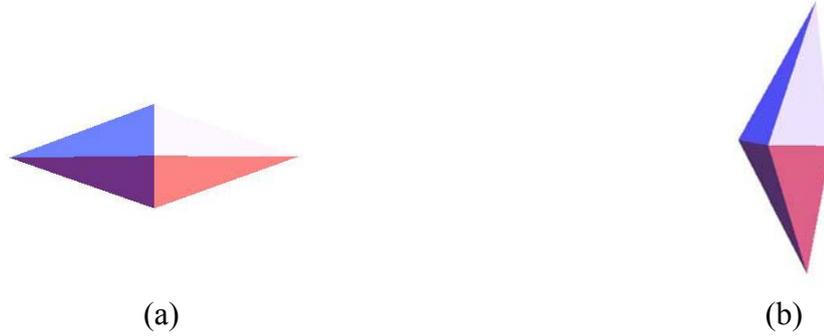

(a)                      (b)

**Figure 9**. (a) A typical cell of the dual polytope of the polytope $O(1,0,1,0)$. (b) Another view.

*5.9. Dual polytope of the polytope* $O(1,1,0,0) = O(\omega_1 + \omega_2)$

This polytope has 2400 vertices and 720 cells. The cells consist of 120 truncated dodecahedra and 600 tetrahedra. The number of cells sharing the same vertex $(1,1,0,0) = \omega_1 + \omega_2$ and the vectors representing their centers are determined as follows:

Truncated dodecahedra : 3 and their centers represented by $\omega_4, r_3 r_4 \omega_4, (r_3 r_4)^2 \omega_4$

Tetrahedron                 : 1 and its center represented by $\lambda \omega_1$.

The scale factor can be determined as $\lambda = \dfrac{\tau + 2}{3\tau + 4}$. This cell is a pyramid with a triangular base with two edge lengths 1.41 and 1.73. The cell is invariant under the dihedral symmetry $D_3 = \langle r_3, r_4 \rangle$. The vertices of the dual polytope can be represented as the union of the following orbits

$$\lambda\, O(1,0,0,0) \oplus O(0,0,0,1). \tag{58}$$

These orbits define two concentric $S^3$ spheres with the ratio of the radii

$$R_1 : R_4 \approx 0.935. \tag{59}$$

The 4 vertices can be expressed in the basis of unit quaternions defined by $p_0 = \dfrac{\omega_1 + \omega_2}{|\omega_1 + \omega_2|}$, $p_1 = e_1 p_0$, $p_2 = e_2 p_0$, $p_3 = e_3 p_0$. Excluding the common components in the direction of $p_0$ they can be determined as follows:

$$\lambda \omega_1 = \lambda[-\tau^2, -1, 0],\ \omega_4 = [1, 0, 3\tau + 1],\ r_3 r_4 \omega_4 = [-1, 2\tau^2, -\tau^2],\ (r_3 r_4)^2 \omega_4 = [\tau^2, -\tau^3, -2\tau]. \tag{60}$$

A plot of this solid is shown in Figure 10.

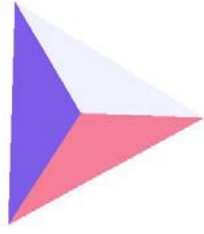

**Figure 10.** A typical cell of the dual polytope of the polytope $O(1,1,0,0)$.

The next four polytopes have dual polytopes whose cells have only one reflection symmetry.

*5.10. Dual polytope of the polytope $O(1,1,1,0) = O(\omega_1 + \omega_2 + \omega_3)$*

This polytope has 7200 vertices and 1920 cells. The cells consist of 120 great rhombicosidodecahedra, 600 truncated tetrahedra and 1200 triangular prisms. The number of cells sharing the same vertex $(1,1,1,0) = \omega_1 + \omega_2 + \omega_3$ and the vectors representing their centers are determined as follows:

Great rhombicosidodecahedra : 2 and their centers represented by $\omega_4, r_4\omega_4$
Truncated tetrahedron : 1 and its center represented by $\lambda\omega_1$
Triangular prism : 1 and its center represented by $\rho\omega_2$.

The scale factors can be determined as $\lambda = \dfrac{5}{5\tau+4}$ and $\rho = \dfrac{5}{3\tau+10}$. This is a solid with 4 vertices possessing a reflection symmetry $r_4$. The faces of the solid consist of two unequal isosceles triangles and two equal scalene triangles. The vertices of the dual polytope can be represented as the union of the following orbits

$$\lambda\, O(1,0,0,0) \oplus \rho\, O(0,1,0,0) \oplus (0,0,0,1). \tag{61}$$

These orbits define three concentric $S^3$ spheres with the ratio of the radii
$$R_1 : R_2 : R_4 \approx 1.0032 : 0.9433 : 1.0000. \tag{62}$$

The 4 vertices can be expressed in the basis of unit quaternions defined by $p_0 = \dfrac{\omega_1 + \omega_2 + \omega_3}{|\omega_1 + \omega_2 + \omega_3|}$, $p_1 = e_1 p_0$, $p_2 = e_2 p_0$, $p_3 = e_3 p_0$. Excluding the common component in the direction of $p_0$ they can be determined as follows:

$$\lambda\omega_1 = \lambda[2\tau^2, \sigma, \tau],\ \rho\omega_2 = \rho[\tau, \tau+2, -\tau],\ \omega_4 = [1,1,4\tau+1],\ r_4\omega_4 = [\tau^2, -2\tau^2, -3\tau]. \tag{63}$$

A plot of this solid is shown in Figure 11.

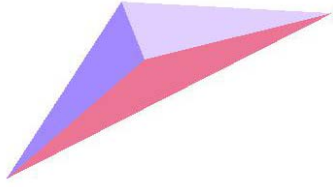

**Figure 11**. A typical cell of the dual polytope of the polytope $O(1,1,1,0)$.

*5.11. Dual polytope of the polytope $O(1,1,0,1) = O(\omega_1 + \omega_2 + \omega_4)$*

This polytope has 7200 vertices and 2640 cells. The cells consist of 120 truncated dodecahedra, 600 cuboctahedra, 720 decagonal prisms and 1200 triangular prisms. The number of cells sharing the same vertex $(1,1,0,1) = \omega_1 + \omega_2 + \omega_4$ and the vectors representing their centers are determined as follows:

Truncated dodecahedron : 1 and its center represented by $\eta\omega_4$
Cuboctahedron : 1 and its center represented by $\lambda\omega_1$
Decagonal prisms : 2 and their centers represented by $\omega_3, r_3\omega_3$
Triangular prism : 1 and its center represented by $\rho\omega_2$.

The scale factors can be determined as $\lambda = \dfrac{6\sigma+7}{4}, \rho = \dfrac{13\sigma+61}{79}$ and $\eta = \dfrac{13\tau+36}{29}$. This is a solid with 5 vertices possessing a reflection symmetry $r_3$. The vertices of the dual polytope can be represented as the union of the following orbits

$$\lambda\, O(1,0,0,0) \oplus \rho\, O(0,1,0,0) \oplus (0,0,1,0) \oplus \eta\, (0,0,0,1). \tag{64}$$

These orbits define four concentric $S^3$ spheres with the ratio of the radii
$$R_4 : R_1 : R_2 : R_3 \approx 1.00 : 1.33 : 1.88 : 1.90. \tag{65}$$

The 4 vertices can be expressed in the basis of unit quaternions defined by $p_0 = \dfrac{\omega_1+\omega_2+\omega_4}{|\omega_1+\omega_2+\omega_4|}, p_1 = e_1 p_0, p_2 = e_2 p_0, p_3 = e_3 p_0$. Excluding the common components in the direction of $p_0$ they can be determined as follows:

$$\begin{aligned}
\lambda\omega_1 &= \frac{\lambda}{2}[2-3\sigma, -2\sigma, -\sigma^2], \quad \rho\omega_2 = \rho[-\tau, \sigma, 2], \quad \eta\omega_4 = \eta[\sigma, 0, -\tau-2], \\
\omega_3 &= [-\sqrt{5}, 2\tau, \sigma], \qquad r_3\omega_3 = [0, 3\sigma-2, -1].
\end{aligned} \tag{66}$$

This is a non-regular pyramid based on a kite base with a $r_3$ symmetry. Plots of this solid are shown in Figure 12.

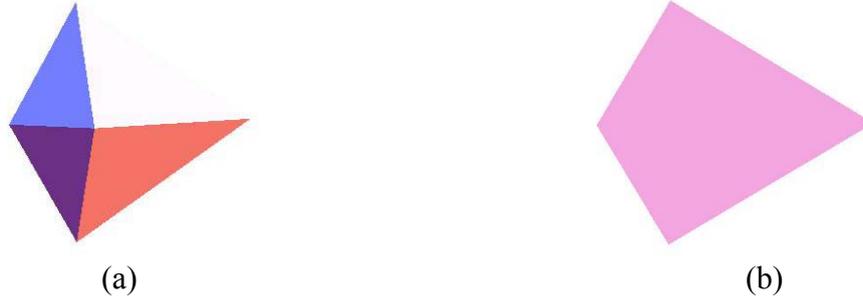

(a) (b)

**Figure 12.** (a) A typical cell of the dual polytope of the polytope $O(1,1,0,1)$. (b) Bottom view.

*5.12. Dual polytope of the polytope* $O(1,0,1,1) = O(\omega_1 + \omega_3 + \omega_4)$

This polytope has 7200 vertices and 2640 cells. The cells consist of 120 small rhombicosidodecahedra, 600 truncated tetrahedra, 720 pentagonal prisms and 1200 hexagonal prisms. The number of cells sharing the same vertex $(1,0,1,1) = \omega_1 + \omega_3 + \omega_4$ and the vectors representing their centers are determined as follows:

Small rhombicosidodecahedra : 1 and its center represented by $\lambda\omega_4$
Truncated tetrahedra : 1 and its center represented by $\rho\omega_1$
Pentagonal prism : 1 and its center represented by $\eta\omega_3$
Hexagonal prisms : 2 and the centers represented by $\omega_2, r_2\omega_2$.

The scale factors can be determined as $\lambda = \dfrac{33\tau + 36}{31}$, $\rho = \dfrac{6\sigma + 27}{19}$ and $\eta = \dfrac{33\tau + 51}{71}$. This is a pyramid with a kite base and a reflection symmetry $r_2$. The vertices of the dual polytope can be represented as the union of the following orbits

$$\rho\, O(1,0,0,0) \oplus O(0,1,0,0) \oplus \eta(0,0,1,0) \oplus \lambda(0,0,0,1). \tag{67}$$

These orbits define four concentric $S^3$ spheres with the ratio of the radii
$$R_1 : R_2 : R_3 : R_4 \approx 1.418 : 1.002 : 1.000 : 1.031. \tag{68}$$

The 5 vertices can be expressed in the basis of unit quaternions defined by $p_0 = \dfrac{\omega_1 + \omega_3 + \omega_4}{|\omega_1 + \omega_3 + \omega_4|}$, $p_1 = e_1 p_0$, $p_2 = e_2 p_0$, $p_3 = e_3 p_0$. Excluding the common component in the direction of $p_0$ they can be determined as follows:

$$\begin{aligned} \rho\omega_1 &= \rho[\sqrt{5}, -1, \sqrt{5}], \quad \eta\omega_3 = \eta[-\tau, \tau, \sigma^2], \quad \lambda\omega_4 = \lambda[\sigma, \sigma, -\tau^2], \\ \omega_2 &= [-1, -\sqrt{5}, 3], \quad r_2\omega_2 = [3, \sqrt{5}, 1]. \end{aligned} \tag{69}$$

Plots of this solid are shown in Figure 13.

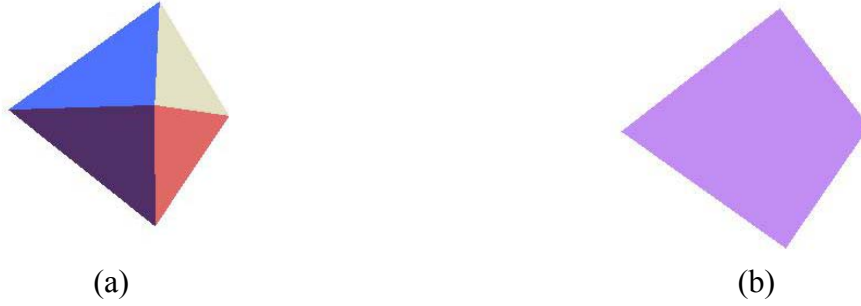

(a) (b)

**Figure 13**. (a) A typical cell of the dual polytope of the polytope $O(1,0,1,1)$. (b) Bottom view.

*5.13. Dual polytope of the polytope $O(0,1,1,1) = O(\omega_2 + \omega_3 + \omega_4)$*

This polytope has 7200 vertices and 1440 cells. The cells consist of 120 truncated icosahedra, 600 truncated octahedra and 720 pentagonal prisms. The number of cells sharing the same vertex $(0,1,1,1) = \omega_2 + \omega_3 + \omega_4$ and the vectors representing their centers are determined as follows:

Truncated icosahedron : 1 and its center represented by $\lambda\omega_4$
Truncated octahedra : 2 and centers represented by $\omega_1, r_1\omega_1$
Pentagonal prism : 1 and its center represented by $\rho\omega_3$.

The scale factors can be determined as $\lambda = \dfrac{6\tau+2}{5}$ and $\rho = \dfrac{6(13\tau+3)}{121}$. This solid consists of two equal scalene triangles and two unequal isosceles triangles as faces possessing a reflection symmetry $r_2$. The vertices of the dual polytope can be represented as the union of the following orbits

$$O(1,0,0,0) \oplus \rho(0,0,1,0) \oplus \lambda(0,0,0,1). \tag{70}$$

These orbits define three concentric $S^3$ spheres with the ratio of the radii

$$R_1 : R_3 : R_4 \approx 1.000 : 0.9906 : 1.0233. \tag{71}$$

The 4 vertices can be expressed in the basis of unit quaternions defined by $p_0 = \dfrac{\omega_2 + \omega_3 + \omega_4}{|\omega_2 + \omega_3 + \omega_4|}$, $p_1 = e_1 p_0$, $p_2 = e_2 p_0$, $p_3 = e_3 p_0$. Excluding the common components in the direction of $p_0$ they can be determined as follows:

$$\omega_1 = [-(2\tau+3), -\sigma, -(\tau+2)],\ r_1\omega_1 = [\tau+3, \tau+2, -(2\tau+1)], \rho\omega_3 = \rho[1, -(\tau+2), 0], \lambda\omega_4 = \lambda[0,1,3\tau].$$
$$\tag{72}$$

A plot of this solid is shown in Figure 14.

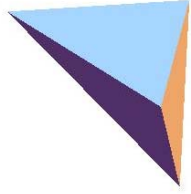

**Figure 14**. A typical cell of the dual polytope of the polytope $O(0,1,1,1)$.

*5.14. Dual polytope of the polytope* $O(1,1,1,1) = O(\omega_1 + \omega_2 + \omega_3 + \omega_4)$

This polytope has 14,400 vertices equal to the order of the group $W(H_4)$ and 2640 cells. The cells consist of 120 great rhombicosidodecahedra, 600 truncated octahedra, 720 decagonal prisms and 1200 hexagonal prisms. The number of cells sharing the same vertex $(1,1,1,1) = \omega_1 + \omega_2 + \omega_3 + \omega_4$ and the vectors representing their centers are determined as follows:

Great rhombicosidodecahedron : 1 and its center represented by $\omega_4$
Truncated octahedron : 1 and its center represented by $\lambda \omega_1$
Hexagonal prism : 1 and its center represented by $\rho \omega_2$
Decagonal prism : 1 and its center represented by $\eta \omega_3$.

The scale factors can be determined as $\lambda = \dfrac{31\sigma + 20}{2}$, $\rho = \dfrac{17\sigma + 30}{57}$ and $\eta = \dfrac{17\sigma + 89}{155}$. This is a solid with 4 vertices possessing no symmetry at all. All its faces are different scalene triangles. The vertices of the dual polytope can be represented as the union of the following orbits

$$\lambda O(1,0,0,0) \oplus \rho(0,1,0,0) \oplus \eta(0,0,1,0) \oplus (0,0,0,1). \tag{73}$$

These orbits define four concentric $S^3$ spheres with the ratio of the radii

$$R_1 : R_2 : R_3 : R_4 \approx 0.9621 : 0.9584 : 09632 : 1.000. \tag{74}$$

The 4 vertices can be expressed in the basis of unit quaternions defined by $p_0 = \dfrac{\omega_1 + \omega_2 + \omega_3 + \omega_4}{|\omega_1 + \omega_2 + \omega_3 + \omega_4|}$, $p_1 = e_1 p_0$, $p_2 = e_2 p_0$, $p_3 = e_3 p_0$. Excluding the common component in the direction of $p_0$ they can be determined as follows:

$$\begin{aligned} \lambda \omega_1 &= \lambda\,[-(5\tau+2), 1, -(3\tau+1)],\ \rho \omega_2 = \rho\,[\tau+1, 3\tau+1, -3(\tau+1)], \\ \eta \omega_3 &= \eta\,[3\tau+1, -2(2\tau+1), \tau], \quad \omega_4 = [\tau, \tau, 4+5\tau]. \end{aligned} \tag{75}$$

A plot of this solid is shown in Figure 15.

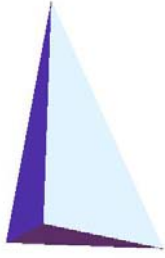

**Figure 15.** A typical cell of the dual polytope of the polytope $O(1,1,1,1)$.

## 6. Conclusion

4D polytopes can be classified with respect to their symmetries represented by the Coxeter groups $W(A_4), W(B_4), W(H_4)$ and $W(F_4)$; the regular and semi regular polytopes are all well known in the mathematical literature. However their constructions in terms of quaternions are recently gaining some interest. Moreover their duals have not been constructed to date. In this paper we have first constructed the group $W(H_4)$ in terms of quaternions and proved that its two 4-dimensional irreducible representations can be constructed with the use of two irreducible representations of the binary icosahedral group represented by the sets of quaternions $I$ and $\tilde{I} = I(\tau \leftrightarrow \sigma)$. We have constructed semi regular polytopes in terms of quaternions and projected them to 3D by decomposing the $W(H_4)$ orbits under the icosahedral group $W(H_3)$. Explicit decomposition of an arbitrary polytope of the group $W(H_4)$ under the group $W(A_4)$ has been made. This decomposition allows the projection of the $W(H_4)$ polytopes into 3D by preserving the tetrahedral symmetry $W(A_3)$. The vertices of the dual polytopes of the semi regular $W(H_4)$ polytopes have been constructed and their cells have been determined.